\documentclass[useAMS,natbib,usegraphicx]{mn2e}
\usepackage{times}
\usepackage{amssymb}
\newcommand{\ttt}[1]{\texttt{#1}}
\def\cgs{{\rm {erg~cm$^{-2}$~s$^{-1}$}}}
\def\nhg{$N_H$(Gal)}
\def\nh{$N_H$}
\def\nhz{$N_H(z=0)$}
\def\cdfn{\ttt{CDFN}}
\def\cdfs{\ttt{CDFS}}

\title[Optically faint hard X-ray sources]
 {X--ray spectral analysis of optically faint sources \\
 in the {\it Chandra} Deep Fields}

\author[F. Civano et al.]
  {Francesca Civano$^{1,2}$\thanks{E-mail: francesca.civano@bo.astro.it (FC); andrea.comastri@bo.astro.it (AC);
  marcella@mpe.mpg.de (MB)}, Andrea Comastri$^2$\footnotemark[1], 
   Marcella Brusa$^{3,2}$\footnotemark[1]\\
  $^1$Dipartimento di Astronomia, Universit\`a di Bologna,
    via Ranzani 1, I--40127 Bologna, Italy\\
  $^2$INAF -- Osservatorio Astronomico di Bologna,  
    via Ranzani 1, I--40127 Bologna, Italy\\
  $^3$Max Planck Institut f\"ur Extraterrestrische Physik,
   Giessenbachstrasse 1, D--85748 Garching, Germany\\
}
\date{Accepted 2005 }

\pubyear{2005}

\def\LaTeX{L\kern-.36em\raise.3ex\hbox{a}\kern-.15em
    T\kern-.1667em\lower.7ex\hbox{E}\kern-.125emX}

\begin{document}


\maketitle

\begin{abstract}
We present the results of a detailed spectral analysis of
optically faint hard X--ray sources in the {\it Chandra}
deep fields selected on the basis of their high X--ray to optical 
flux ratio (X/O). The stacked spectra of high X/O sources 
in both {\it Chandra} deep fields, fitted with a single power--law model, 
are much harder than the spectrum of the X--ray background (XRB).
The average slope is also insensitive to the 2--8 keV 
flux, being approximately constant around $\Gamma~\simeq$~1 over 
more than two decades, strongly indicating that high X/O 
sources represent the most obscured component of the XRB.
For about half of the sample, a redshift estimate (in most of the
cases a photometric redshift) is available from the literature. 
Individual fits of a few of the brightest objects and of 
stacked spectra in different redshift bins imply column 
densities in the range 10$^{22-23.5}$~cm$^{-2}$.
A trend of increasing absorption towards higher redshifts 
is suggested.

\end{abstract}


\section{Introduction}

Thanks to the deep surveys carried out by {\it Chandra} and 
XMM--{\it Newton} X--ray satellites, the hard (2--8 keV) 
X--ray sky is now probed down to a flux limit of about 
$2\times~10^{-16}$~\cgs\ where a fraction as large as 
80\% of the diffuse XRB is resolved into discrete sources 
(Mushotzky et al. 2000; Hasinger et al. 2001; Rosati et al. 2002; 
Alexander et al. 2003).

The outcomes of these surveys confirm the basic predictions of 
AGN synthesis models: the integrated emission of obscured and unobscured AGN
folded with an appropriate evolving luminosity function is responsible 
of the spectral intensity of the hard {\tt XRB}
(Setti \& Woltjer 1989; Comastri et al. 1995; 
Gilli, Salvati \& Hasinger 2001).
The expected source counts at both hard and soft X--ray energies
are also in agreement with the observed logN--logS 
(Moretti et al. 2003); moreover, the observed average spectrum of 
the resolved source population almost exactly matches that of the
XRB ($\langle \Gamma \rangle \sim1.4$; Tozzi et al. 2001; Nandra et al. 2004).

Massive campaigns of optical and near--infrared
follow-up observations have reached a high
completeness level. 
If, on the one hand, the observed redshift and absorption 
distribution are only poorly reproduced by AGN synthesis models 
for the X--ray background,  
calling for some revisions of the underlying assumptions, 
on the other hand it has been pointed 
out that the observed redshift distribution could be significantly
biased against the optically faintest, presumably obscured high redshift 
objects (Fiore et al. 2003; Treister et al. 2004).
A solid estimate of the high luminosity, high--redshift tail of the
AGN luminosity function is a key parameter to understand  
the evolution of accretion powered sources and their present
day quiescent massive black holes which appear to be ubiquitous 
in local galaxies (see, e.g., Ferrarese \& Merritt 2000).

It is well known that the X--ray to optical flux ratio
(hereinafter X/O, see next section for a definition) could provide a first, 
rough indication of the source classification (Maccacaro et al. 1988). 
The majority of X--ray selected, relatively unobscured quasars and Seyfert 
galaxies are characterized by $-1 < X/O < 1$, while 
obscured AGN have, on average, higher X--ray to optical 
flux ratios. Combining the X--ray and optical photometry 
of several hundreds of sources detected by both {\it Chandra} 
and XMM--{\it Newton} over a broad range of fluxes, about
20\% of hard X--ray selected objects are characterized 
by high X/O values ($>$ 1) and their fraction does not depend
upon the X--ray flux (Brusa 2004). 
The optical magnitudes 
of high X/O sources detected at relatively bright 
X--ray fluxes are of the order of  R$\simeq$24 and thus 
accessible to spectroscopy with large telescopes.
Well defined samples of high X/O sources detected in the
{\tt HELLAS2XMM} surveys at X--ray fluxes 
brighter than $\sim$10$^{-14}$\cgs\ have been the subject
of intensive FORS/VLT optical spectroscopy (Fiore et al. 2003) and
near--infrared $K_s$ ISAAC/VLT photometry (Mignoli et al. 2004).
The results clearly indicate that the majority of the targets
are high--redshift ($z>$ 1), luminous, X--ray obscured AGN 
(Perola et al. 2004). The X/O selection is thus highly efficient in sampling 
high redshift obscured quasars (see also Brusa et al. 2004).

 Independent arguments suggesting obscured accretion at 
high redshifts as a likely
explanation of high X/O ratios have been discussed by Comastri, Brusa and 
Mignoli (2003).
Moving the Spectral Energy Distribution (SED) of an X--ray absorbed
AGN to progressively higher redshifts the K--corrections
in the optical and X--ray band work in the opposite direction.
The ratio between the optical to X--ray optical depth, in the observer frame,
scales roughly as $(1+z)^{3.6}$, 
because dust extinction increases in
the UV while X-ray absorption strongly decreases going toward high
energies.  The net result is that in the presence of an absorbing
screen the observed optical flux of a high-z AGN can be strongly
reduced, and the observed magnitudes are mainly due to starlight in
the host galaxies. Conversely, the hard  X--ray flux is  much
less affected.  
The observed high values of the X/O are therefore at least
qualitatively consistent with those expected by a
population of high redshift, absorbed AGN with X--ray column densities
in the range N$_{H}$=10$^{22}$--$10^{24}$ cm$^{-2}$.

At the limits of deep {\it Chandra} surveys, high X/O sources
are characterized by extremely faint optical magnitudes and 
spectroscopic identification is not feasible even with the largest 
telescopes.
In this framework a detailed investigation of their X--ray 
properties may provide useful information on the nature
of this important component of the X--ray source population. 
We present and discuss in the following the results of detailed X--ray 
spectral analysis, including the
search for iron $K\alpha$ lines, combined
with the available multiwavelength information 
for a large sample of high X/O sources in the  
{\it Chandra} Deep Field North (Alexander et al. 2003) and the 
{\it Chandra} Deep Field South (Giacconi et al. 2002; 
hereinafter \cdfn\ and \cdfs). 
A cosmological model with $H_0$ = 70 km s$^{-1}$ Mpc$^{-1}$,
$\Omega_m$ = 0.3 and $\Omega_{\Lambda}= 0.7$ is assumed.

\section{Sample selection}
\label{sec:sample}
\begin{figure}
\includegraphics[angle=0,width=85mm]{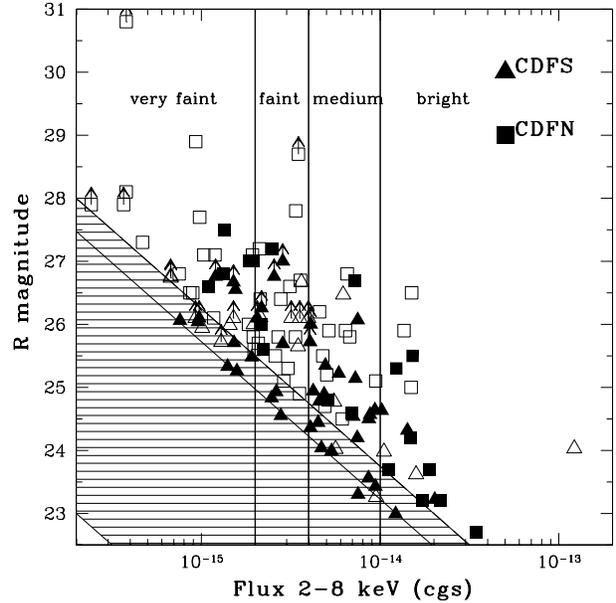}
\caption{The R--band magnitude vs. the 2--8 keV flux for the 127
high X/O sources selected from the \cdfn\ (squares) and \cdfs\ (triangles) 
samples. Filled symbols are objects with
redshift identification. The dashed area represents the locus occupied by 
known AGN (e.g. quasars, Seyferts, emission--line galaxies) along the 
correlation log(X/O)=0$\pm$1. Two upper boundaries at log(X/O)=1 are reported
as we used two different constants for the \cdfn\ and the \cdfs\ (see Sect.~2 
for details). The vertical lines are the boundaries of the 4 flux
bins discussed in Section 4.1. }
\label{fig:one}
\end{figure}

We have selected 127 hard X--ray sources in the
{\it Chandra} Deep Fields (63 in the \cdfn\ and
64 in the \cdfs)  with an X--ray to optical 
flux ratio $\log{(f_X/f_R)}>1$ defined as:

\begin{equation}
\label{eqz:unica}
X/O=\log{\frac{f_X}{f_R}}=\log{f_X}+C+\frac{R}{2.5}
\end{equation}

where $f_X$ is the X--ray flux (\cgs) in a given energy range,
$f_R$ is the R--band flux computed by converting the R--band magnitudes into
monochromatic fluxes (at $\lambda$=6500$\AA$) and then 
multiplying them by the width of the R filter (Zombeck 1990), 
C is a constant which depends on the specific filter used in the optical
observations. In order to be consistent with the published optical catalogues we used 
C=5.5 for the \cdfn\ and C=5.71 for the \cdfs.
The hard band (2--8 keV) fluxes and the basic X--ray source 
properties (positions, counts, exposure times) have
been compiled from the Alexander et al. (2003) X--ray source 
catalogue for \cdfn\ and \cdfs.
The R--band magnitudes along with optical identification and 
classification (when available) have been retrieved from 
Barger et al. (2003) for the \cdfn, Giacconi
et al. (2002), Szokoly et al. (2004) and Zheng et al. (2004) for the \cdfs. 
Figure~\ref{fig:one} shows the R--band magnitude versus the 2--8 keV
flux for the 127 high X/O sources discussed in this work.

For each source we report in Table~1 and 2 the X--ray
identification number (col.~1), the X--ray coordinates (col.~2 and 3),
the hard X--ray flux (col.~4), the optical R--band magnitude (col.~5), 
the magnitude reference for \cdfs\  (col.~6) and the X/O
from equation~\ref{eqz:unica} (col.~7 and 6, respectively).
Sources undetected in the R band are reported as lower limits in 
Tables~1--2 and in these cases the X/O ratio has to be 
considered a lower limit. All the high X/O objects
represent $\sim$ 23\% of the hard X--ray sources revealed in both deep fields.
 
\begin{table*}
\centering
\begin{minipage}{120 mm}
\caption{Properties of high X/O sources in the \cdfs.}
\begin{scriptsize}
\begin{tabular}{ccccccccc}
\hline
\hline
XID & RA & DEC & f$_X$ $^a$ & R & Ref. $^b$ & X/O & $z$ & $z$ type $^c$ \\
\hline 
  3  &3:33:05.85&  -27:46:50.2 &      3.18 &   $>$26.10 &   G  & $>$1.65	&  ... &  $^+$       \\
  8  &3:33:01.44&  -27:41:42.0 &      9.40 &    23.43 &   S  & 1.06&  0.99   & phot (0.9)    \\	
 20a &3:32:44.47&  -27:49:40.2 &      5.34 &    23.99 &   S  & 1.03&  1.016 &  spec        \\
 23  &3:32:41.85&  -27:43:59.9 &      2.62 &    24.93 &   S  & 1.10&  0.73  & phot (0.5)\\	
 25  &3:32:40.84&  -27:55:46.6 &      9.34 &    24.64 &   S  & 1.54&  2.26  & phot (0.5)        \\
 26  &3:32:39.73&  -27:46:11.2 &      2.78 &    24.55 &   S  & 0.97$^\dagger$&  1.624 & phot (0.5)        \\
 27  &3:32:39.68&  -27:48:50.7 &      7.06 &    24.54 &   S  & 1.37&  3.064 & spec         \\
 31  &3:32:37.77&  -27:52:12.4 &      8.78 &    24.57 &   S  & 1.48&  1.603 & spec         \\
 35  &3:32:34.37&  -27:39:13.1 &      14.2 &    24.32 &   S  & 1.59&  1.51  & spec       \\
 45  &3:32:25.68&  -27:43:05.8 &      4.94 &    25.35 &   S  & 1.54&  2.291 & spec         \\
 48  &3:32:24.84&  -27:56:00.0 &      4.86 &    24.89 &   G  & 1.35&  0.841 & phot$^d$\\
 51a &3:32:17.18&  -27:52:20.9 &      20.2 &    23.22 &   S  & 1.30&  1.097 & spec         \\
 54  &3:32:14.57&  -27:54:21.6 &      2.85 &    25.69 &   S  & 1.44&  2.561 & spec         \\
 57  &3:32:12.95&  -27:52:36.7 &      4.69 &    24.04 &   S  & 1.00&  2.562 & spec         \\
 58  &3:32:11.77&  -27:46:28.2 &      1.96 &    26.00 &   S  & 1.40&  ...   &  $^{*}$           \\ 
 59  &3:32:11.41&  -27:52:13.5 &      6.21 &    26.47 &   S  & 2.09&  ...  &   $^{++}$          \\ 
 61  &3:32:10.50&  -27:43:09.0 &      10.5 &    23.98 &   S  & 1.32&  ...  &   $^{*}$          \\
 64  &3:32:08.00&  -27:46:57.2 &      5.52 &    24.77 &   S  & 1.36&  ...  &   $^{+}$          \\
 65  &3:32:03.90&  -27:53:28.9 &      2.17 &   $>$26.26 &   S  & $>$1.55&  1.10  & phot (0.5)        \\
 67  &3:32:02.46&  -27:46:00.3 &      8.61 &    23.56 &   S  & 1.07&  1.616 & spec         \\
 69a &3:32:01.43&  -27:41:38.6 &      5.63 &    24.02 &   S  & 1.07& ...   &  $^+$           \\
 70a &3:32:01.42&  -27:46:47.1 &      15.9 &    23.62 &   S  & 1.36& ...    &  $^+$         \\
 72a &3:31:58.29&  -27:50:41.6 &      7.48 &    26.07 &   S  & 2.01&  1.99  &  phot (0.5)       \\
 76  &3:31:52.50&  -27:50:17.5 &      8.64 &    24.50 &   S  & 1.45&   2.394&  spec        \\
 79  &3:32:38.03&  -27:46:26.2 &      1.55 &    26.55 &   S  & 1.52&   1.91 &  phot (0.5)       \\
 81  &3:32:25.96&  -27:45:14.3 &      0.95 &    26.04 &   S  & 1.11&  2.59  &  phot (0.5)       \\
 82a &3:32:14.87&  -27:51:03.8 &      1.51 &    $>$26.1 &   S  & $>$1.33&  ...   &  $^{*}$          \\
 99  &3:32:05.17&  -27:53:54.7 &      5.87 &    25.22 &   S  & 1.57&  0.79  &  phot (0.5)       \\
108a &3:32:05.77&  -27:44:46.6 &      1.01 &    25.94 &   S  & 1.09& ...  &    $^{*}$         \\ 
112 &3:31:51.98&  -27:53:26.7 &      2.47 &   24.83  &   S  & 1.03&   2.942&  spec        \\
133  &3:32:02.52&  -27:44:29.8 &      0.67 &   $>$26.74 &   G  & $>$1.23&  ...   &   $^{++}$          \\
145  &3:32:22.54&  -27:46:03.9 &      4.56 &    24.78 &   S  & 1.28&  1.50  &  phot (0.5)       \\
146  &3:32:47.05&  -27:53:33.3 &      2.55 &   $>$26.76 &   G  & $>$1.82& 2.67	&  phot (0.5)       \\
147  &3:32:46.35&  -27:46:31.9 &      7.26 &    25.14 &   S  & 1.63&  0.99  &  phot (0.5)       \\
148  &3:32:35.23&  -27:53:17.8 &      2.85 &   $>$27    &   S  & $>$1.96&  1.74  &  phot (0.5)       \\
153  &3:32:18.34&  -27:50:55.2 &      7.45 &    24.20 &   S  & 1.26&  1.536 &  spec        \\
159  &3:32:50.23&  -27:52:51.7 &      7.53 &    23.30 &   S  & 0.90$^\dagger$&  3.3	&  phot (0.5)       \\
179  &3:31:49.49&  -27:50:34.0 &      0.97 &   $>$26.10 &   G  & $>$1.14&  2.73  &  phot (0.5)       \\
201a &3:32:39.06&  -27:44:39.1 &      2.14 &    26.11 &   S  & 1.48& ... 	&    $^{*}$          \\
205a &3:32:17.11&  -27:41:36.6 &      1.44 &    25.99 &   S  & 1.26&  ...  &    $^{++}$         \\
207a &3:32:07.96&  -27:37:35.1 &      122.0 &    24.03 &   S  & 2.41&  ...  &     $^+$        \\
209  &3:31:47.30&  -27:53:13.3 &      10.2 &    24.63 &   S  & 1.57&1.32	&   phot (0.5)      \\
213a &3:32:00.56&  -27:53:53.0 &      3.47 &    25.65 &   S  & 1.51& ...   &   $^{++}$           \\
219  &3:31:50.43&  -27:51:51.8 &      3.96 &   $>$26.10 &   G  & $>$1.75& 	...  &    $^{*}$     \\
227  &3:32:05.35&  -27:46:44.1 &      3.62 &    26.67 &   S  & 1.94&  ...   &   $^{*}$         \\
243  &3:32:08.39&  -27:40:47.0 &      4.05 &   $>$25.72 &   G  & $>$1.61&  2.50$^{e}$  & phot (0.3)       \\
253  &3:32:20.07&  -27:44:47.0 &      4.22 &    24.94 &   S  & 1.31&  1.89  &  spec      \\
256  &3:32:43.05&  -27:48:45.0 &      4.07 &    24.36 &   S  & 1.06&  1.53  &  phot (0.5)       \\
259  &3:32:06.13&  -27:49:27.7 &      4.50 &    24.44 &   S  & 1.14& 1.76	&  phot (0.5)       \\
261  &3:31:57.03&  -27:51:08.6 &      3.55 &   $>$26.10 &   G  & $>$1.70& ... &   $^{**}$        \\
263b &3:32:18.89&  -27:51:35.4 &      1.40 &    25.33 &   S  & 0.99$^\dagger$&3.66	&  spec       \\
265  &3:32:33.27&  -27:42:36.1 &      4.08 &   $>$26.00 &   G  & $>$1.72& 1.16	&  phot (0.5)       \\
501  &3:33:10.18&  -27:48:41.8 &      12.2 &    22.99 &   S  & 0.99$^\dagger$&  0.81  &  phot (0.6)     \\
503  &3:33:07.61&  -27:51:26.6 &      9.48 &    23.26 &   S  & 0.99$^\dagger$&  ...  & $^+$           \\
505a &3:33:04.83&  -27:47:31.9 &      1.91 &    25.48 &   S  & 1.18&   0.981&  phot$^d$\\    
506  &3:33:02.97&  -27:51:46.4 &      0.93 &   $>$26.10 &   G  & $>$1.12&  ...   &    $^{*}$         \\
508  &3:32:51.64&  -27:52:12.8 &      1.20 &   $>$26.76 &   G  & $>$1.49&  2.5	&  phot (0.5)       \\
513a &3:32:34.03&  -27:48:59.9 &      0.76 &    26.06 &   S  & 1.01&  3.53  &  phot (0.5)       \\
515  &3:32:32.17&  -27:46:51.4 &      1.51 &   $>$26.67 &   G  & $>$1.56&  2.19  &  phot (0.5)       \\
518  &3:32:26.75&  -27:46:04.3 &      0.67 &   $>$26.76 &   G  & $>$1.21&  ...  &    $^{++}$         \\
524  &3:32:19.96&  -27:42:43.2 &      1.29 &   $>$25.72 &   G  & $>$1.11&  ...  &    $^{*}$         \\
533  &3:32:13.90&  -27:56:00.1 &      2.05 &   $>$26.10 &   G  & $>$1.46&  0.54  &  phot (0.5)       \\
595  &3:32:15.75&  -27:39:53.8 &      1.52 &    25.71 &   S  & 1.18& 0.36	&  phot (0.5)       \\
606a &3:32:24.97&  -27:50:08.2 &      1.58 &    25.26 &   S  & 1.01&  1.037 & spec         \\
\hline
\hline
\end{tabular}
\end{scriptsize}
\end{minipage}
\begin{minipage}{155 mm}
$^a$ In units of $10^{-15}$~erg~cm$^{-2}$~s$^{-1}$.
$^b$ G = Giacconi et al. 2002, S = Szokoly et al. 2004.
$^c$ The photo--z quality flag (QF) from Zheng et al. (2004) is reported in
brackets;
QF=0.3 means that the redshift is from the BPZ code; QF=0.5 means that
two indipendent codes (BPZ and HyperZ) return consistent values; QF=0.6 means that HyperZ code 
returns the same value of COMBO-17 catalogue; QF=0.9 means that the value returned by BPZ and HyperZ 
is confirmed in the COMBO-17 catalogue. 
$^d$  Photometric redshift from COMBO-17 survey catalogue (Wolf et al. 2004).
$^e$ Although QF=0.3, the detection of a K$\alpha$ line (Gilli, private communication) at the 
best--fit photo--z makes us confident on this value.
$^\dagger$ The 5 sources with X/O slightly lower than 1 would have 
X/O $>$ 1 if the X--ray flux quoted in Giacconi et al. (2001) 
is considered.
$^+$ Sources with QF$\leq$0.4 in Zheng et al. (2004).
$^{++}$ QF=0.5 but with a large error on redshift estimate ($\Delta z \geq$1.0).
$^{*}$ The optical counterpart presented in Zheng et al. (2004) lies at $>2"$ from the 
Alexander et al. (2003) position.
$^{**}$ Source 261 is not present in Zheng et al. (2004). 
\end{minipage}
\end{table*}
\begin{table*}
\centering
\begin{minipage}{100 mm}
\caption{Properties of high X/O sources in the \cdfn.}
\begin{scriptsize}
\begin{tabular}{cccccccc}
\hline
\hline
XID & RA & DEC & f$_X$ $^a$& R & X/O & $z$ & $z$ type $^b$ \\
\hline 							    	
7      &  12:35:21.71    &  +62:15:01.6    &   2.11	     &  27.2   & 1.704	  &    ...      &  ...  \\
9      &  12:35:24.92    &  +62:15:24.8    &   4.77      &  25.5   & 1.378	  &      ...    & ...   \\
10     &  12:35:28.77    &  +62:14:27.8    &   4.58      &  26.2       & 1.640	  &   ...       & ...   \\
12     &  12:35:29.45    &  +62:18:22.8    &   3.36      &  25.8       & 1.346	  &    ...      & ...   \\
28     &  12:35:46.07    &  +62:15:59.9    &   2.48      &  27.2       & 1.774	  &     1.93    &  phot\\
29     &  12:35:46.25    &  +62:17:29.8    &   1.93      &  27.1       & 1.625	  &   ...    	  & ...   \\
36     &  12:35:50.42    &  +62:18:08.6    &  12.3	     &  25.3       & 1.71	  &   0.52	 & K$\alpha$\\
38     &  12:35:51.75    &  +62:17:57.1    &  15.0	     &  26.5       & 2.276	  &  ...  	   & ...   \\
42     &  12:35:54.11    &  +62:20:12.3    &   0.86      &  26.5       & 1.035	  &  ...  	   & ...   \\
48     &  12:35:56.14    &  +62:12:19.2    &  18.8	     &  23.7       & 1.25	  &    1.13&	K$\alpha$\\
71     &  12:36:05.62    &  +62:06:54.0    &  1.95	     &  25.8       & 1.11	   &  ...           &  ...   \\
73     &  12:36:05.83    &  +62:08:38.0    &   2.78      &  26.4      & 1.50	   &  ...           &  ...   \\
91     &  12:36:11.40    &  +62:21:49.9    &   9.44      &  25.1       & 1.51	   &  ...           &  ...   \\
98     &  12:36:13.02    &  +62:12:24.1    &   1.97      &  25.6    & 1.03	   &  ...           &  ...   \\
92     &  12:36:11.80    &  +62:10:14.5    &   2.69      &  25.8    & 1.24	   &   ...          &  ...   \\
98     &  12:36:13.02    &  +62:12:24.1    &   1.97      &  25.6    & 1.03	   &   ...          &  ...   \\
100    &  12:36:14.14    &  +62:10:17.7    &   0.93      &  28.9    & 2.028	   &   ...          &  ...   \\
102    &  12:36:14.45    &  +62:10:45.0    &   3.54      &  24.9   & 1.00	   &   ...          &  ...  \\
107    &  12:36:15.83    &  +62:15:15.5    &   1.85      &  27.0    & 1.567	  &  2.06	  & phot\\
108    &  12:36:16.03    &  +62:11:07.7    &   15.2	     &  25.5    & 1.88	  &  1.25	  & phot\\
129    &  12:36:21.94    &  +62:16:03.8    &   1.32      &  26.8    & 1.34	  &  2.26	  & phot\\
134    &  12:36:22.65    &  +62:10:28.5    &   2.89      &  25.1    & 1.00	  &  ...    	    &  ...   \\
140    &  12:36:23.66    &  +62:10:08.7    &   0.47     &  27.3    & 1.09	  &  ...    	    &  ...   \\
146    &  12:36:25.33    &  +62:17:37.7    &   2.21      &  25.6    & 1.084	  &  1.28  & phot\\
151    &  12:36:27.53    &  +62:12:18.0    &   0.24      & $>$27.9    & $>$1.04	  &    ...          &  ...   \\
154    &  12:36:28.65    &  +62:21:39.5    &   14.9	     &  25      & 1.673	  &    ...          &  ...   \\
156    &  12:36:28.78    &  +62:11:40.0    &   0.38	     & $>$30.8  & $>$2.39	  &       ...       &  ...   \\
165    &  12:36:30.15    &  +62:26:20.1    &   11.1	     &  23.7       & 1.025	  &   1.46	 & phot\\
186    &  12:36:34.71    &  +62:04:37.9    &   2.08      &  25.7       & 1.098	  &    ...  	    &  ...   \\
195    &  12:36:36.85    &  +62:22:27.4    &   2.13      &  26.4    & 1.38	  &   ...           &  ...   \\
196    &  12:36:36.90    &  +62:13:20.0    &   0.38      &  28.1    & 1.32	  &   ...           &  ...   \\
204    &  12:36:38.94    &  +62:10:41.5    &   0.37     & $>$27.9    & $>$1.22	  &     ...         &  ...   \\
232    &  12:36:44.95    &  +62:26:51.0    &   4.91     &  24.7    & 1.07	  &    ...          &  ...   \\
246    &  12:36:47.94    &  +62:10:19.9    &   1.94      &  27.0    & 1.58	  &  2.52	 & phot \\
250    &  12:36:48.28    &  +62:14:56.2    &   1.34      &  27.5    & 1.62	  &  2.09	& phot \\
253    &  12:36:48.73    &  +62:21:53.9    &   2.59      &  25.5    & 1.11	  &   ...  	&  ...   \\
259    &  12:36:49.66    &  +62:07:38.3    &   21.8	     &  23.2    & 1.11	  &	 1.609   & spec\\
290    &  12:36:56.56    &  +62:15:13.1    &   0.75     &  26.8    & 1.09	  &  ...          &  ...   \\
318    &  12:37:01.76    &  +62:07:20.8    &   0.98     &  27.7    & 1.57	  & ...           &  ...   \\
321    &  12:37:02.43    &  +62:19:26.1    &   3.05      &  25.3    & 1.10	  &  ...          &  ...   \\
334    &  12:37:04.87    &  +62:16:01.6    &   6.40      &  25.9   & 1.66	  &  ...          &  ...   \\
335    &  12:37:05.12    &  +62:16:34.8    &   1.03      &  27.1    & 1.35	  & ...           &  ...   \\
336    &  12:37:05.31    &  +62:24:54.8    &   3.12      &  26.6    & 1.63	  &  ...          &  ...   \\
350    &  12:37:07.70    &  +62:05:34.6    &   1.83      &  26         & 1.16	  &  ...          &  ...   \\
357    &  12:37:09.40    &  +62:22:14.4    &   6.94      &  24.6       & 1.18	  &   1.43  &  phot\\
374    &  12:37:13.84    &  +62:18:26.2    &   1.19      &  27.1    & 1.41	  & 	 ...  &   ...  \\
382    &  12:37:15.19    &  +62:02:31.0    &   6.76      &  25.8    & 1.64	  &  ...  	&  ...   \\
390    &  12:37:16.65    &  +62:17:33.3    &   17.3      &  23.2    & 1.01	  &    1.146    & spec\\
400    &  12:37:20.26    &  +62:07:26.7    &   0.89      &  26.5     & 1.05	  &  ...         &  ...   \\
406    &  12:37:22.44    &  +62:05:36.1    &   3.37      &  27.8     & 2.14	  &  ...         &   ...  \\
412    &  12:37:24.00    &  +62:13:04.3    &   6.13      &  24.5      & 1.08	  & ...  	   &  ...   \\
413    &  12:37:24.29    &  +62:13:59.7    &   5.09      &  24.8      & 1.13      &    0.474	  & spec\\
444    &  12:37:36.85    &  +62:14:29.2    &   5.16      &  25.9       & 1.57	  &    ...       &   ...  \\
445    &  12:37:37.04    &  +62:18:34.4    &   3.58      &  26.7     & 1.73	  &  ...         &  ...   \\
452    &  12:37:39.46    &  +62:22:39.2    &   3.49      & $>$28.7   & $>$2.52	  &	 ...    &  ...    \\
456    &  12:37:41.13    &  +62:10:47.9    &   1.09      &  26.6       & 1.77	  &1.60   & phot  \\
463    &  12:37:45.02    &  +62:07:18.9    &   7.20      &  26.7       & 2.03	  & 2.50 &  phot\\
465    &  12:37:46.78    &  +62:07:12.6    &   1.17      &  26.1       & 1.00	  &  ...     &  ...  \\
470    &  12:37:50.22    &  +62:13:59.3    &   2.17      &  26.0       & 1.23	  &    0.23	   & spec\\
490    &  12:38:10.56    &  +62:17:29.6    &  14.8	     &  24.2    & 1.35	  &   1.02	      & phot\\
495    &  12:38:22.30    &  +62:14:16.7    &  34.4	     &  22.7   & 1.11	  &    0.986   &spec \\
498    &  12:38:23.66    &  +62:09:42.4    &   5.03      &  25.2       & 1.28	  &  ...             &  ...  \\
497    &  12:38:23.21    &  +62:15:18.3    &   6.54      &  26.8       & 2.03	  &  ...             &  ...  \\
501    &  12:38:34.29    &  +62:14:40.9    &   13.6      &  25.9       & 1.99	  &  ...             &  ...  \\
\hline
\hline
\end{tabular}
\end{scriptsize}
\end{minipage}
\newpage
\begin{minipage}{100 mm}
$^a$ In units of $10^{-15}$~erg~cm$^{-2}$~s$^{-1}$.
$^b$ Redshift type: spectroscopic, photometric or from the K$\alpha$ iron line.

\vspace{1.5 cm}
\end{minipage}
\end{table*}

\section{Data reduction}
The \cdfs\ 1 Megasecond dataset is the result of the 
co-addition of 11 individual {\it Chandra} ACIS--I 
exposures whose aimpoints are at a few arcsec from each
other. The total area covered is $\sim$392~arcmin$^2$; 
this field was selected in a patch of the southern sky 
(nominal aim point $\alpha_{2000}=03^{h}32^{m}28^s.0$ e $\delta_{2
000}=-27^\circ48^{\prime}30^{\prime\prime}$) characterized by 
a very low Galactic neutral hydrogen column density 
(8$\times$10$^{19}$~cm$^{-2}$) and by the lack of bright stars 
(Rosati et al. 2002).\\
The \cdfn\ has been observed by about 2 Megaseconds 
over a period of 27 month, and is the result of
the co-addition of 20 individual {\it Chandra} 
ACIS-I exposures centered (nominal aim point $\alpha_{2000}=12^h36^m49^s.4$ 
 $\delta_{2000}=62^\circ12^{\prime}58^{\prime
\prime}$) close to the Hubble Deep Field North 
(HDF--N, Williams et al. 1996), the most studied region of the sky with a 
low Galactic neutral hydrogen column density 
(1.6$\times$10$^{20}$~cm$^{-2}$, Stark et al. 1992). 
Due to the different pointings, required to keep the HDF--N near the aim
point and the {\it Chandra} roll angle 
constraints, the area covered is  447.8 arcmin$^2$.  

The X--ray data have been retrieved from the public archive 
and processed with standard tools making use of the calibrations 
associated with the {\tt CIAO}\footnote{http://cxc.harvard.edu/ciao/}
software (version 3.0). All the observations of both fields have been registered on the deepest one 
and have been aligned using {\tt align\_evt}\footnote{http://cxc.harvard.edu/cal/ASPECT/align\_evt/}.

Since the main aim of the present work concerns the 
average spectral properties of relatively faint sources,
the issue of co-adding spectra obtained from observations
performed with different aim points and roll angles
has to be treated appropriately.
In principle
for each pointing and each object the source 
and background spectra, along with the detector 
position--dependent response matrix and effective area
should be computed and then summed together weighting 
for the exposure time of each observation (``standard'' procedure).
In practice, such an approach results to be extremely time consuming.
In order to compute, from the total merged event file, all the relevant
files needed for the spectral analysis 
an alternative approach has been devised. This approach is based on two {\tt CIAO} tools 
{\tt mkwarf} and {\tt mkrmf} 
\footnote{http://asc.harvard.edu/ciao/threads/wresp\_multiple\_sources/}
developed to extract spectral data from multiple regions
either from the same or different observations.
More specifically, the spectrum and relative background for 
a sample of sources are extracted from 
a stack of regions whose shape and size are function 
of the off--axis angle.  The weighted detector position dependent  
spectral response and effective area 
are computed taking into account the shape and area of the extraction 
regions and the number of counts of each individual region 
in the stack.
In order to test whether these tools could be applied to multiple extraction 
regions in co-added observations,
we proceeded as follows.


We considered several different samples of N sources, 
randomly distributed over the detector in both  
the \cdfn\ and \cdfs\ merged event files.
The size and shape of extraction regions were chosen according to 
the number of counts and source off--axis angle in each exposure
and tuned to enclose the 90\% of the PSF at 6.4 keV. 
This energy has been chosen since our sources have been selected 
in the hard band ($\S$\ref{sec:sample}).
Several background spectra were extracted from a stack
of regions nearby each source and varied in size and shape. 
The corresponding response {\tt RMF} and effective area {\tt ARF} file 
have been computed from the merged event files 
making use of {\tt mkwarf} and {\tt mkrmf}. 
The ancillary response files are corrected for the well known degradation of the
{\tt ACIS} quantum efficiency using the latest version
of the {\tt ACISABS} tool \footnote{http://chandra.ledas.ac.uk/ciao3.0/threads/apply\_acisabs/}. 


In order to check the reliability of our procedure with respect to the standard one,
we have computed the average stacked spectrum of 30 X--ray sources
in the \cdfn\ and 35 in the \cdfs\ spectroscopically identified 
as broad--line AGN by Barger et al. (2003) and Szokoly et al. (2004),
respectively. The best--fit parameters of a single absorbed power law are:
$\Gamma$=1.73$\pm$0.03,  
\nhz=7$\pm$1$\times$10$^{20}$~cm$^{-2}$ in the \cdfn\ 
and $\Gamma$=1.82$\pm$0.04,  
\nhz=4.7$\pm$1.44$\times$10$^{20}$~cm$^{-2}$ in the \cdfs\ , 
slightly flatter ($\Delta\Gamma\sim$0.1-0.2) than the ``canonical'' 
value (e.g. $\Gamma$=1.9--2.0; Nandra \& Pounds 1994), but fully consistent
with the fact that no reflection component has been included in the spectral fitting. 
Moreover, in the \cdfn\ our results 
are in excellent agreement with those reported by 
Bauer et al. (2002) for a sample of broad--line AGN 
largely overlapping with ours. 
Given that the Bauer et al. results have been obtained 
applying a rather different (IDL based) data reduction 
procedure (ACIS-Extract),
we are confident of the reliability of our method.

\section{Spectral analysis} 
 The stacked spectra have been extracted in the 0.7--7 keV band, to minimize 
residual calibration uncertainties of the low energy instrumental
response, and rebinned to have at least 20 counts per bin.
X--ray spectral fitting was performed using XSPEC 
(version 11.3.0; Arnaud 1996); 
errors are reported at the 90\%
confidence level for one interesting parameter ($\Delta\chi^2$=2.71).
In all the fits, a ``negative''  edge ($\tau = -0.17$) at 2.07 keV
has been included to take into account calibration uncertainties 
around the instrumental Iridium edge (Vikhlinin et al. 2005).

\subsection{Average spectra}
\label{secAS}
The stacked spectra of the high X/O sources in both
the \cdfn\ (about 30000 net counts) and \cdfs\ 
(about 19500 net counts) were fitted with an absorbed power--law model 
with absorption fixed at the Galactic value (hereinafter MODEL 1). 

\begin{figure*}
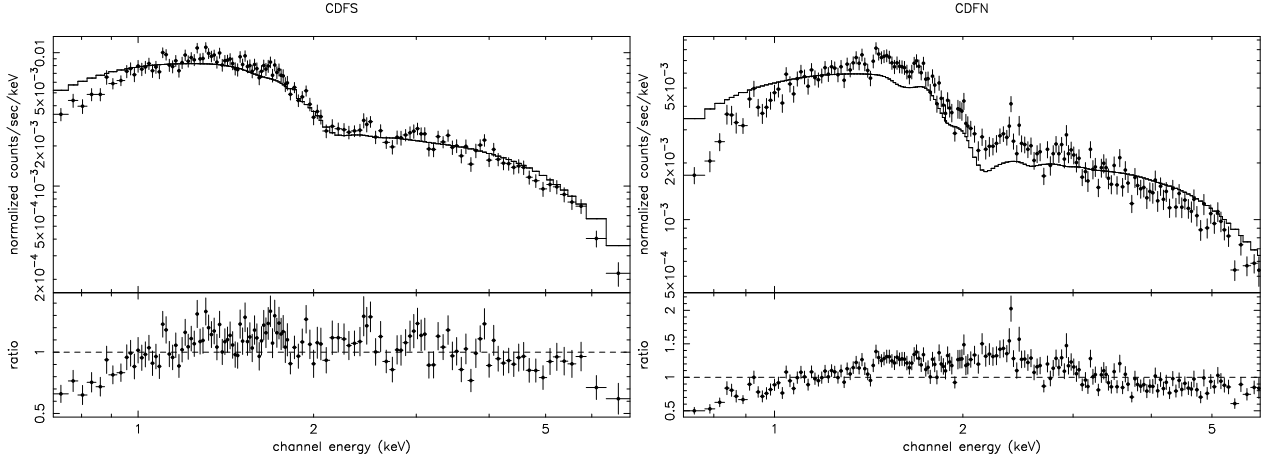

\includegraphics[width=60mm,angle=-90]{ME1314rv_fig_2_1.ps}
\includegraphics[width=60mm,angle=-90]{ME1314rv_fig_2_2.ps}
\caption{Stacked spectra of all the high X/O sources in \cdfs\ (64 sources) and \cdfn\
(63 sources), fitted in the 0.7-7 keV energy band with a power--law spectrum (MODEL 1). The best--fit spectral slopes are 
reported in Table~3. The solid line is the best-fit model, while the lower panel shows the data-to-model ratio.}
\label{fig:two}
\end{figure*}

The resulting best--fit slopes are $\Gamma$(\cdfn)=0.77$\pm$0.02 
and $\Gamma$(\cdfs)=0.97$\pm$0.03. 
Although the residuals (Fig. \ref{fig:two}) 
indicate a more complex curved shape, these  
extremely hard spectra strongly suggest that high X/O sources 
represent the most obscured component of the so far
resolved XRB. Even more obscured 
possibly Compton thick AGN, yet undetected in deep 
surveys, might be responsible of the XRB at higher energies
(Worsley et al. 2004).
The measured values are significantly flatter 
than the average spectrum of the total \cdfs\ sample  
in the 1--10 keV range 
($\langle \Gamma \rangle$=1.375$\pm$0.015; Rosati et al. 2002) as well as 
of the average XRB slope ($\Gamma \simeq$ 1.4) in the same energy range.

\begin{figure*}
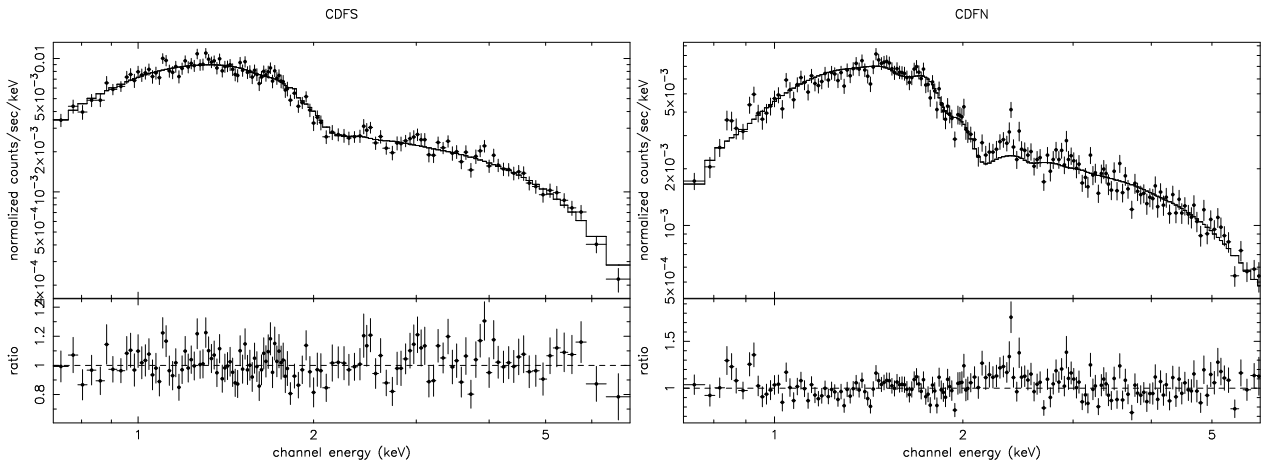

\includegraphics[width=60mm,angle=-90]{ME1314rv_fig_3_1.ps}
\includegraphics[width=60mm,angle=-90]{ME1314rv_fig_3_2.ps}
\caption{Stacked spectra of all the high X/O sources selected in \cdfs\ (64 sources) and \cdfn\
(63 sources), fitted in the 0.7-7 keV energy band with an absorbed power--law spectrum (MODEL 2). The best--fit column densities and spectral slopes are
reported in Table~3. The solid line is the best-fit model , while the lower panel shows the data-to-model ratio.}
\label{fig:three}
\end{figure*}

Leaving the absorption free to vary (hereinafter MODEL 2) 
the quality of the fit improves significantly (Fig. \ref{fig:three}).
The best--fit slopes and column densities are:
$\Gamma$=1.34$\pm$0.04, 
\nhz=4.1$\pm$0.3$\times$10$^{21}$~cm$^{-2}$ for the \cdfn\ 
and $\Gamma$=1.34$\pm$0.03,  
\nhz=2.4$\pm$0.4$\times$10$^{21}$~cm$^{-2}$ for the \cdfs.
Given that the stacked objects are likely to be distributed
over a range of redshifts and that no redshift dependence has been considered, 
the \nh\ values have to be considered as lower limits to 
any absorption component.
For the same reason the intrinsic power--law slope might well be steeper,
the observed value being the result of a distribution of column densities 
convolved with redshift.
 
\begin{figure*}
\includegraphics[angle=0,width=100 mm]{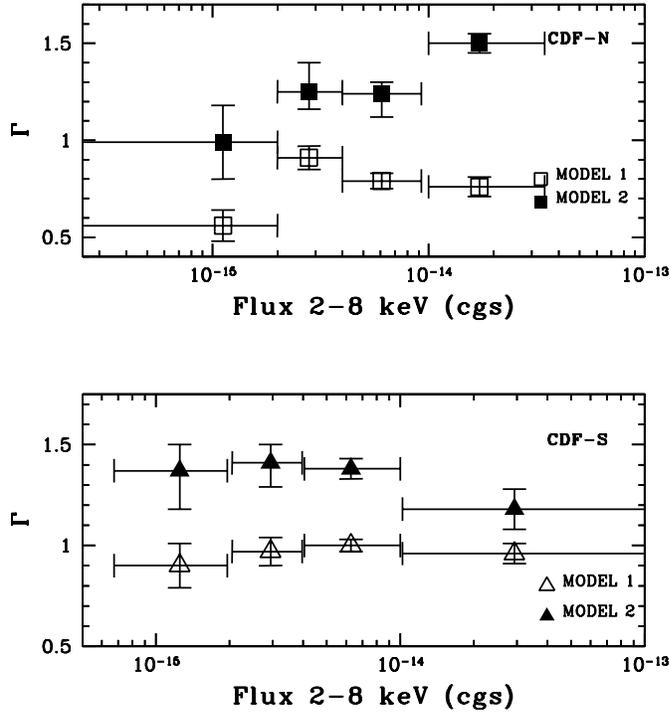}
\caption{Spectral index $\Gamma$ vs. hard X-ray flux obtained with MODEL 1 (open symbols) and MODEL 2 (filled symbols) for 
the 4 subsamples of \cdfn\ (squares, top panel) and \cdfs\ (triangles, bottom panel). The errors on the slope are at the 90\% confidence
level; the horizontal bars represent the bin width. }
\label{fig:four}
\end{figure*}
\begin{figure*}
\includegraphics[angle=0,width=100 mm]{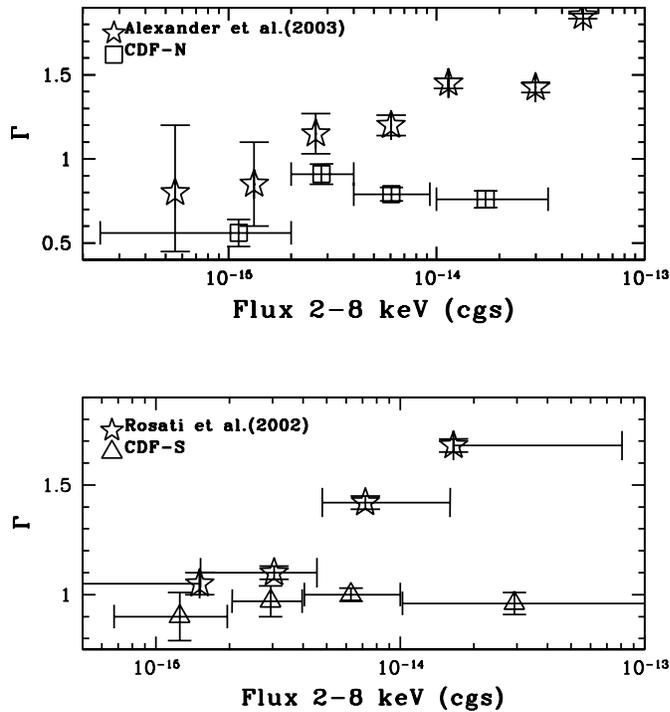}
\caption{Top panel: Spectral index $\Gamma$ vs. hard X-ray flux obtained with MODEL 1 for high X/O \cdfn\ subsamples (squares) 
compared with the values of $\Gamma$ as obtained from band ratios for 
the entire \cdfn\ sample (stars, Alexander et al. 2001,
2003). Bottom panel: Spectral index 
$\Gamma$ vs hard X-ray flux obtained with MODEL 1 for high X/O \cdfs\ subsamples (triangles) 
compared with the values obtained for all \cdfs\ hard sample (stars, Rosati et al. 2002) }
\label{fig:five}
\end{figure*}
\begin{table*}
\centering
\begin{minipage}{100mm}
\caption{Spectral fit parameters obtained with MODEL 1 and MODEL 2.}
\begin{tabular}{c c c c c c c}
\hline
\hline
&Sample$^a$ & Model & $\Gamma$ & \nhz$^b$  & $\chi^2$/d.o.f. \\   
\hline
\cdfs\  & all (64) & MODEL 1 &0.97$\pm$0.03 &Galactic$^c$    & 429.2/261               \\ 
	&  & MODEL 2 & 1.34$\pm$0.03 &0.24$\pm$0.04	& 255.8/260                \\ 
	
  & A (7)&MODEL 1&0.96$\pm$0.05		& Galactic	 &133.6/109	\\ 
  &  &MODEL 2&1.18$\pm$0.1	   &0.15$\pm$0.05		    &115.9/108	   \\ 
 
  &B (25)&MODEL 1&1.00$\pm$0.03	   &Galactic			    &368.3/254	   \\ 
  &&MODEL 2&1.38$\pm$0.05 &0.25$\pm$0.04	    &259.1/253	   \\ 
 
  &C (14)&MODEL 1&0.97$\pm$0.07	   &Galactic			    &137.2/125	   \\ 
  &&MODEL 2&1.41$^{+0.09}_{-0.12}$&0.27$^{+0.09}_{-0.06}$	    &107.3/124	   \\ 
 
  &D (18)&MODEL 1&0.90$\pm$0.11	   &Galactic			    &105.9/99	   \\ 
  &&MODEL 2&1.37$^{+0.13}_{-0.19}$&0.31$^{+0.17}_{-0.11}$	    &92.3/98	   \\  
\hline							
\cdfn\ & all (63) & MODEL 1 & 0.77$\pm$0.02  & Galactic$^d$      &   1137.7/374       \\ 
	&  & MODEL 2 & 1.34$\pm$0.04 & 0.41$\pm$0.03  &   473.2/373          \\ 
 
  &A (11)&MODEL 1&0.76$\pm$0.05	   &Galactic		    &1091.0/257	   \\ 
  & &MODEL 2&1.50$\pm$0.05	   &0.53$\pm$0.03	    &361.7/256 $^e$    \\ 
 
  &B (13) &MODEL 1&0.79$\pm$0.04	   &Galactic		    &279.0/168 	   \\ 
  &&MODEL 2&1.24$^{+0.06}_{-0.12}$&0.30$\pm$0.05	    &196.7/167	   \\ 
 
  &C (17) &MODEL 1&0.91$\pm$0.06	   &Galactic		    &180.2/151	   \\  
  &&MODEL 2&1.25$^{+0.15}_{-0.09}$&0.23$^{+0.08}_{-0.04}$  &146.0/150	   \\ 
 
  &D (22)&MODEL 1&0.56$\pm$0.08	   &Galactic		    &133.4/109	   \\ 
  &&MODEL 2&0.99$\pm$0.19 & 0.33$^{+0.14}_{-0.10}$  & 113.6/108  	   \\ 
\hline
\hline
\end{tabular}	
\end{minipage}
\begin{minipage}{100mm}
$^a$ The number of sources in each bin is reported in brackets.
$^b$ Units of $10^{22}$ cm$^{-2}$.
$^c$\nhg (\cdfs) = 8$\times$10$^{19}$ cm$^{-2}$.
$^d$\nhg (\cdfn) =1.6$\times$10$^{20}$ cm$^{-2}$.
$^e$ The quality of the fit is not good due to features around 3 keV (see Sect. 4.2).
\end{minipage}
\end{table*}

The accumulated source counts in the stacked spectra are sufficient
to further investigate the spectral properties of 
high X/O sources and, in particular, to 
search for the progressive hardening of the mean spectral slope towards 
faint fluxes; this hardening has been clearly established 
both in the \cdfs\ (Rosati et al. 2002) 
and \cdfn\ (Alexander et al. 2003) total samples.

The high X/O sources of our sample were divided in four 
2--8 keV flux intervals (see Fig. \ref{fig:one}).
The bin width was optimized to keep the total number 
of counts in each bin adequate to perform 
spectral analysis. 
The flux intervals were defined as follows: bright (A, $>$ 10$^{-14}$~\cgs), medium 
(B, 4$\times$10$^{-15}$--10$^{-14}$~\cgs), 
faint (C, 2$\times$10$^{-15}$--4$\times$10$^{-15}$~\cgs) and 
very faint (D, $<$2$\times$10$^{-15}$~\cgs).
Stacked spectra in each subsample were extracted following the procedure
described in Sect.~3 and fitted with both MODEL 1 
and MODEL 2: the best--fit parameters are reported 
in Table~3, and shown in Figure \ref{fig:four}.
There is no clear evidence of a spectral flattening 
as a function of the X--ray flux, with a 
possible exception for the \cdfn\ sources fitted with MODEL 2. 
The average spectrum of high X/O sources remains constant 
with a very hard slope ($\Gamma \lesssim$ 1)  over 
about two decades of X--ray flux and definitely flatter 
than that measured for the entire \cdfs\ and \cdfn\ samples 
down to comparable limiting fluxes (see Fig. \ref{fig:five}).
The difference is more pronounced ($\Delta\Gamma\sim$0.7) at  
fluxes brighter than  $\sim$ 5$\times$ 10$^{-15}$ \cgs, 
which are dominated by unobscured quasars with a soft X--ray spectrum, 
decreasing to $\Delta\Gamma\sim$0.15 at fainter fluxes, where 
the contribution of obscured sources with X/O~$<$~1 to the 
total sample is much higher. \\ 
A very flat 2--10 keV spectrum may also be the signature of
a strong Compton reflection component and is expected for 
column densities in the Compton--thick regime (\nh $\gtrsim 2 \times$ 
10$^{24}$ cm$^{-2}$). Spectral fits performed with a pure
reflection spectrum (model {\tt pexrav} in {\tt XSPEC}) do 
not provide a statistically acceptable description of the 
data leaving significant residuals at low energy.
The most straightforward explanation of the observed flat 
slope in the stacked spectra is in terms of the superposition
of different photoelectric cut-offs due to Compton--thin absorption
likely spread over a broad range of redshifts.

\subsection{Redshift determination through K$\alpha$ iron line}
\label{sec:line}

The residuals of a single power--law fit to the average spectrum
of the \cdfn\ bright subsample (A) are highly indicative of the presence 
of a line--like feature. 
Given that the iron line at 6.4 keV is by far the strongest emission 
feature in the X--ray band, it can be used as a redshift indicator.
The 11 sources in subsample A have a counting statistic
which is such to allow for individual spectral analysis: spectra were
extracted for these sources and fitted with MODEL 2.
A clear excess suggesting the presence of an iron--line 
has been found in 4 of them (XID 36, 48, 259, 390). 
The Fe$K\alpha$ line of source 390 
has been already extensively discussed by 
Comastri, Brusa \& Civano (2004). 
A spectroscopic redshift (z$_{spec}$=1.609) has been measured 
by  Barger et al. (2003) for source 259; this value is fully consistent 
with that inferred from the iron $K\alpha$ line ($z_{X-ray}$~=1.60$\pm$~0.05).

\begin{figure*}
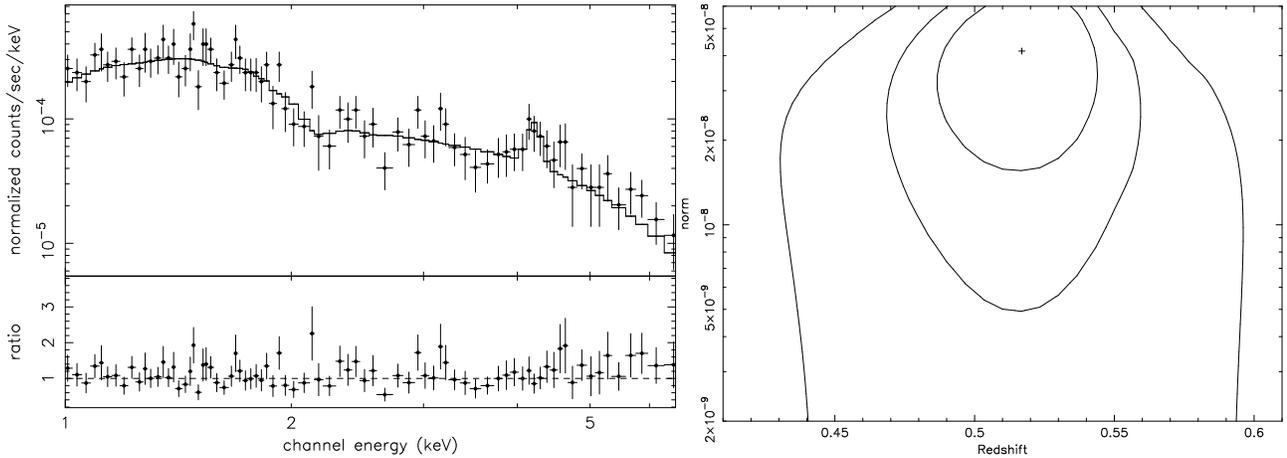

\includegraphics[width=60mm,angle=-90]{ME1314rv_fig_6_1.ps}
\includegraphics[width=60mm,angle=-90]{ME1314rv_fig_6_2.ps}
\caption{Left panel: the X--ray spectrum of source 36 (\cdfn), fitted with an absorbed power--law plus a FeK$\alpha$ line (the best 
fit column density and spectral slope are \nh$~=~1.20\pm0.15~\times~10^{22}~cm^{-2}$ and $\Gamma = 1.56\pm0.07$) 
and residuals of the fit. Right panel: 68, 90, 99\% confidence contours of the
redshift versus line intensity. } 
\label{fig:six}
\end{figure*}

\begin{figure*}
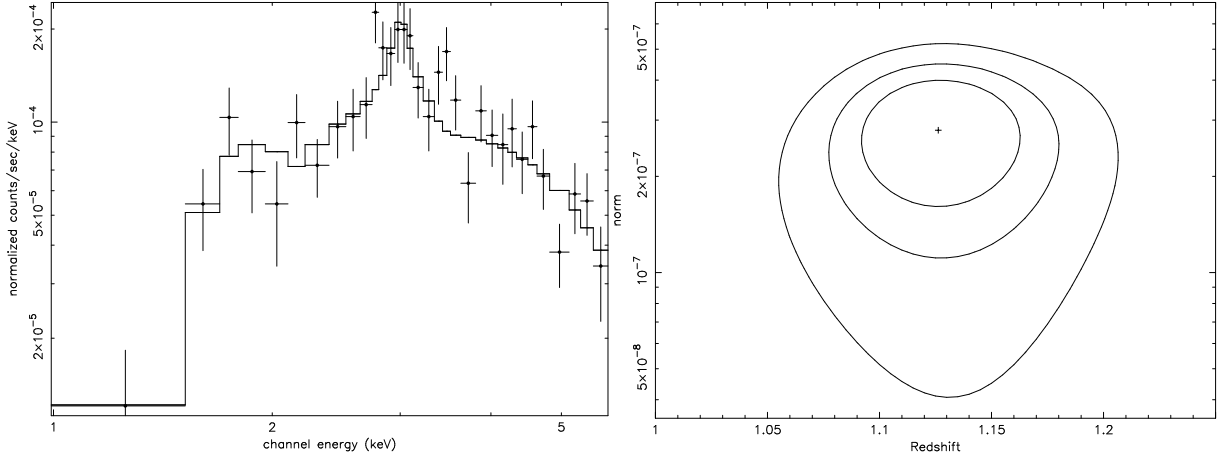

\includegraphics[width=60mm,angle=-90]{ME1314rv_fig_7_1.ps}
\includegraphics[width=60mm,angle=-90]{ME1314rv_fig_7_2.ps}
\caption{Left panel: the X--ray spectrum of source 48 (\cdfn), fitted with an absorbed power--law plus a FeK$\alpha$ line (the 
best--fit column density and spectral slope are \nh$~=~3.2\pm0.3~\times~10^{23}~cm^{-2}$ and $\Gamma = 1.88\pm0.07$). 
Right panel: 68, 90, 99\% confidence contours of the
redshift versus the line intensity. } 
\label{fig:seven}
\end{figure*}

The addition of a narrow Gaussian line due to cold iron 
($E_{rest}$~=~6.4 keV) in MODEL 2 improves 
the fit by $\Delta\chi^2 \simeq$~9 and $\simeq$~15 for source 36 and 48, 
respectively.
The observed line energy corresponds to an X--ray redshift 
of $z=0.52^{+0.03}_{-0.05}$ and $z=1.13\pm0.05$.
In figure \ref{fig:six} and \ref{fig:seven} the spectra and the 
confidence contours of redshift versus line intensity are shown.
The rest frame equivalent widths are 293$\pm$204 eV and  400$^{+201}_{-195}$ eV, respectively.
Assuming the redshift estimated by the iron line, the inferred column densities  
are \nh$~=~1.20\pm0.15~\times~10^{22}$~cm$^{-2}$ 
and \nh$~=~3.2\pm0.3~\times~10^{23}$~cm$^{-2}$, respectively.
The absorption corrected 2--10 keV luminosities 
are 1.3$\times$10$^{43}$~erg~s$^{-1}$ and 
1.1$\times$10$^{44}$~erg~s$^{-1}$ respectively. 
The measured values of intrinsic absorption and luminosity classify
source 48 as a type 2 quasar candidate.
%
\subsection{Average spectra as a function of redshift}
\label{sec:redshift}
\begin{table*}
\centering
\vspace{3cm}
\begin{minipage}{60mm}
\caption{Redshift values for each redshift bin: in col.~1 the mean 
redshift weighted by the counts of the sources in each bin, in col.~2 and 3 the maximum and
minimum redshift of the sources in each bin. }
\begin{tabular}{c c c c}
\hline
\hline
&mean $z$ & min $z$ & max $z$ \\
\hline
\cdfs\ &0.80	 &0.36   &0.99 \\ 
	&1.15	 &1.016  &1.32 \\
	&1.68	 &1.51   &1.99 \\
	&2.46    &2.19   &2.942\\
	&3.12	 &3.064  &3.66 \\
\cdfn\ &0.49	 &0.474  &0.52 \\
	&1.09    &0.986  &1.146\\
	&1.46    &1.25   &1.609\\
	&2.27	 &1.93   &2.52 \\
\hline
\hline
\end{tabular}	
\end{minipage}
\end{table*}
\begin{table*}
\centering
\begin{minipage}{100mm}
\caption{Spectral fit parameters of the spectra in redshift bins obtained with a power law model plus intrinsic absorption at
the mean redshift (the two entries for each 
bin correspond to a fit with $\Gamma$ free to vary and to a fit with $\Gamma$=1.8).}
\begin{tabular}{c c c c c c c}
\hline
\hline
& $z^a$  &number       &  $\Gamma$ & \nh$^b$ (1) &  $\chi^2$/d.o.f. \\   
&    &of sources&&&\\
\hline
\cdfs\  &0.80  &9&  1.29$\pm$0.09 & 0.44$^{+0.31}_{-0.24}$      & 117.0/111  \\ 
      &	&  &  1.8& 1.32$\pm$0.23                         & 144.0/112  \\ 
      
      &1.15  & 6& 0.90$^{+0.08}_{-0.11}$  & 2.56$^{+1.19}_{-1.11}$    & 76.7/74  \\ 
      &	&  & 1.8  & 8.39$\pm$0.8                      &147.1/75   \\ 
      
      & 1.68  &12 & 1.41$\pm$0.06 &   1.51$^{+0.58}_{-0.48}$   &  184.7/161     \\ 
      & 	& & 1.8 &  3.29$^{+0.41}_{-0.39}$             & 218.3/162      \\ 

	& 2.46  & 12 & 1.35$^{+0.11}_{-0.13}$  & 12.98$^{+3.23}_{-2.52}$    & 97.1/102    \\ 
      	& 	&  & 1.8  & 19.98$^{+2.51}_{-2.26}$  &  114.1/103    \\ 

	  & 3.12  & 4 & 1.20$^{+0.12}_{-0.14}$  & 8.86$^{+4.93}_{-3.57}$    & 44.6/50    \\ 
	 & 	 &  & 1.8  & 20.62$^{+4.20}_{-3.65}$  &  65.0/51    \\ 
\hline							
\cdfn\       &  0.49  & 3&  1.23$\pm$0.12  &  1.39$^{+0.48}_{-0.26}$&  83.6/80    \\ 
	     &  	 &  &1.8&  2.48$^{+0.28}_{-0.26}$&  103.4/81    \\  
	
	    &  1.09 &  4 &1.18$\pm$0.06 &  2.74$^{+0.43}_{-0.34}$&286.6/190$^c$    \\ 
 	     & 	 & &1.8&  5.54$\pm$0.36 & 403.4/191$^c$     \\  
 	
	&  1.46 & 6  & 1.58$\pm$0.05&  5.62$^{+0.60}_{-0.50}$&   278.7/215  \\ 
     	&  	 &    &1.8&  6.92$^{+0.38}_{-0.39}$ & 295.9/216    \\
   
	 &  2.27 & 6 &0.98$\pm$0.15	&   4.32$^{+2.75}_{-2.00}$&  72.6/63    \\
       & 	 &  &1.8&   12.91$^{+3.06}_{-2.55}$& 104.9/64   \\ 
\hline
\hline
\end{tabular}	
\end{minipage}
\begin{minipage}{100mm}
$^a$ Mean value weighted by the counts of the sources in each bin.
$^b$ Units of $10^{22}$ cm$^{-2}$.  
$^c$ The statistical quality of the fit is bad due to the presence of a strong Fe k$\alpha$ iron line at E=2.9 keV 
(Comastri, Brusa \& Civano 2004) and to another iron line at E$\simeq$3.0 keV.
\vspace{3cm}
\end{minipage}
\end{table*}

The hypothesis that strong obscuration is ubiquitous among
high X/O sources could be quantified once a redshift estimate 
is available. 
In the \cdfs, after a careful check of the optical X--ray associations 
which were found to be discrepant between the published X-ray and optical
catalogues (Giacconi et al. 2002; Alexander et al. 2003; Szokoly et al. 2004;
Zheng et al. 2004),
we have considered in addition to the 15 spectroscopic redshifts, 
28 photo--z.
Most of them (25) have been selected from 
the Zheng et al. (2004) compilation requiring a quality flag $\geq$0.5 
\footnote{A quality flag 0.5 means that 
two independent codes 
(the BPZ--Bayesian Photometric Redshift estimation, Benitez 2000; 
HyperZ, Bolzonella et al. 2000) return consistent values.},
two from the COMBO--17 survey (Wolf et al. 2004) and for 
one further object, though the photo--z quality flag is 0.3,  
the detection of an iron line at the same
redshift (Gilli, private communication) makes us confident on the redshift 
estimate. Five sources satisfying the quality flag 
criterion have been excluded due to the extremely large errors
quoted by Zheng et al. (2004). In the \cdfn\ we consider all the 
available redshifts, 5 spectroscopic plus 12 reliable photo--z 
(according to Barger et al. 2003) computed with the BPZ code. 
Not surprisingly, the sources without a redshift measurement are, 
on average, optically fainter (most of them with R$>$25).
For two sources in the \cdfn\ a reliable redshift estimate has been obtained
directly from the X--ray data thanks to the detection of
a K$\alpha$ line (see Sect. \ref{sec:line}).

All the redshifts considered for the spectral analysis are reported in 
column 8 of Table~1 and in column 7 of 
Table~2, along with a flag in column 9 and 8, respectively
(phot = photometric, spec = spectroscopic, K$\alpha$ = iron line).

The 62 sources with spectroscopic (20), photometric (40) and X--ray
(2) redshift were subdivided in 9 redshift bins, 5 for \cdfs\ and 4 
for \cdfn\ (see Table~4 for details). Each bin is centered at the 
redshift value obtained
by weighting the source counts for a given interval. 

\begin{figure*}
\includegraphics[angle=0,width=110mm]{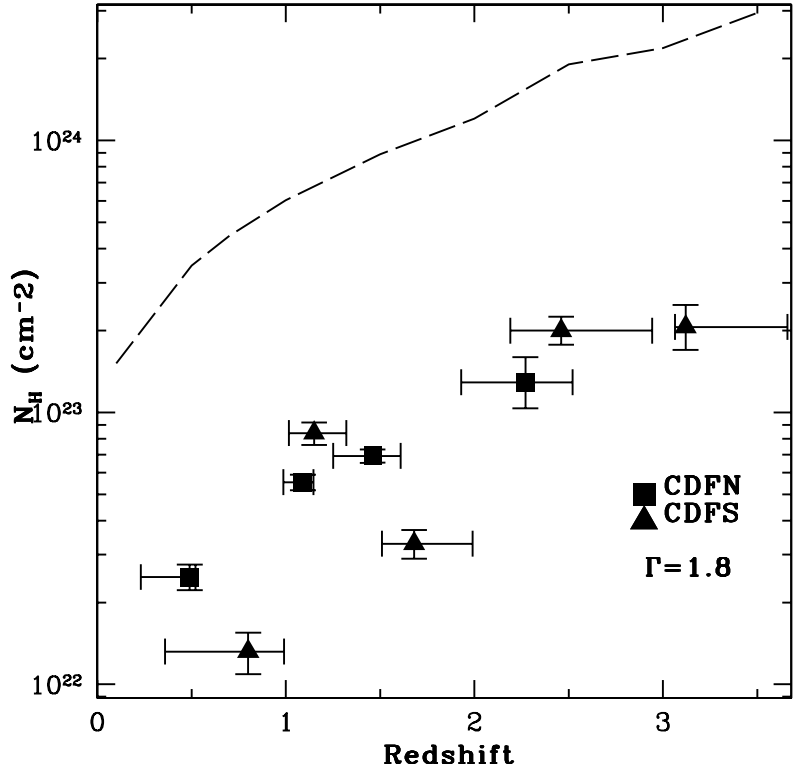} 
\caption{Intrinsic \nh\ rest frame distribution for the stacked spectra in different redshift bins. 
Symbols as in figure \ref{fig:four}. The dashed line represents the detection 
limit obtained by the simulations described in Section 4.3.}
\label{fig:eight}
\end{figure*}

Stacked spectra of the sources in each redshift bin were extracted
and fitted with a a power--law model plus intrinsic absorption 
at the mean redshift.
At first, \nh\ and $\Gamma$ were free to vary, then 
the power--law slope has been fixed to $\Gamma$=1.8
in order to better constrain the intrinsic absorption column density. 
The 95\% confidence level errors on the photo--z of eight \cdfs\ sources 
is larger than the bin width.
We have verified that excluding these sources 
does not significantly modify the results and thus all of them were 
included and assigned at the redshift bin corresponding to the best 
fit photo--z. The best--fit \nh\ values are reported 
in Table~5, and plotted in Figure \ref{fig:eight}. 

The results in both fields unambiguously confirm that high X/O sources 
are highly obscured with rest frame 
column densities  larger than 10$^{22}$~cm$^{-2}$.
The average value of the column density at $z \simeq 3$ 
is of the order of 2$\times$10$^{23}$~cm$^{-2}$. 
As an independent check, the individual spectra of the brightest 
objects, spread over the entire redshift range and with enough
counts ($>$ 200) to perform moderate--quality spectral 
analysis, have been fitted with a $\Gamma$=1.8 power--law
plus rest--frame absorption. In all the cases the fitted 
\nh\ is consistent with the value obtained for the stacked spectrum 
in the corresponding redshift bin.\\ 
\begin{figure*}
\includegraphics[width=60mm]{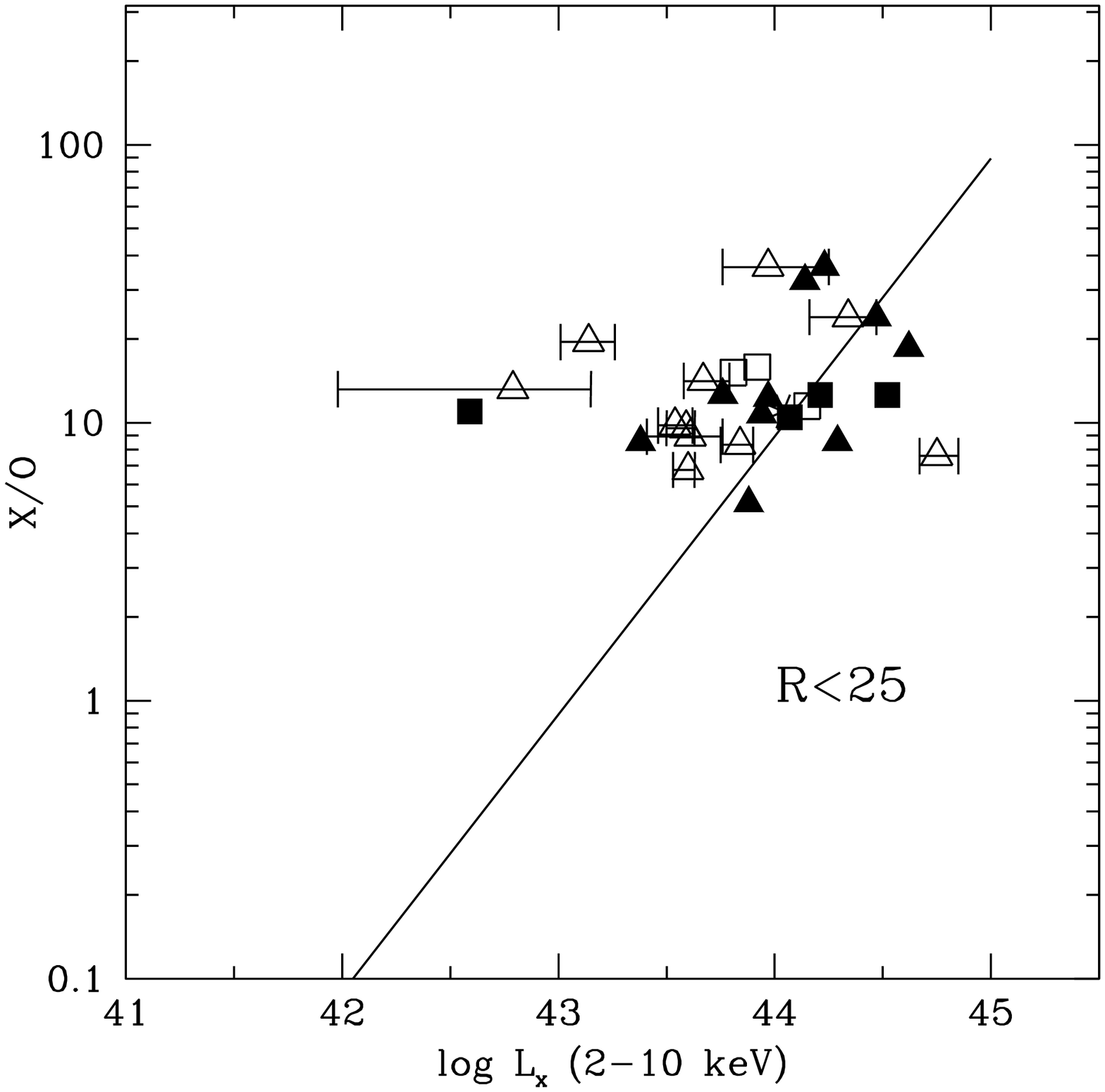}
\includegraphics[width=60mm]{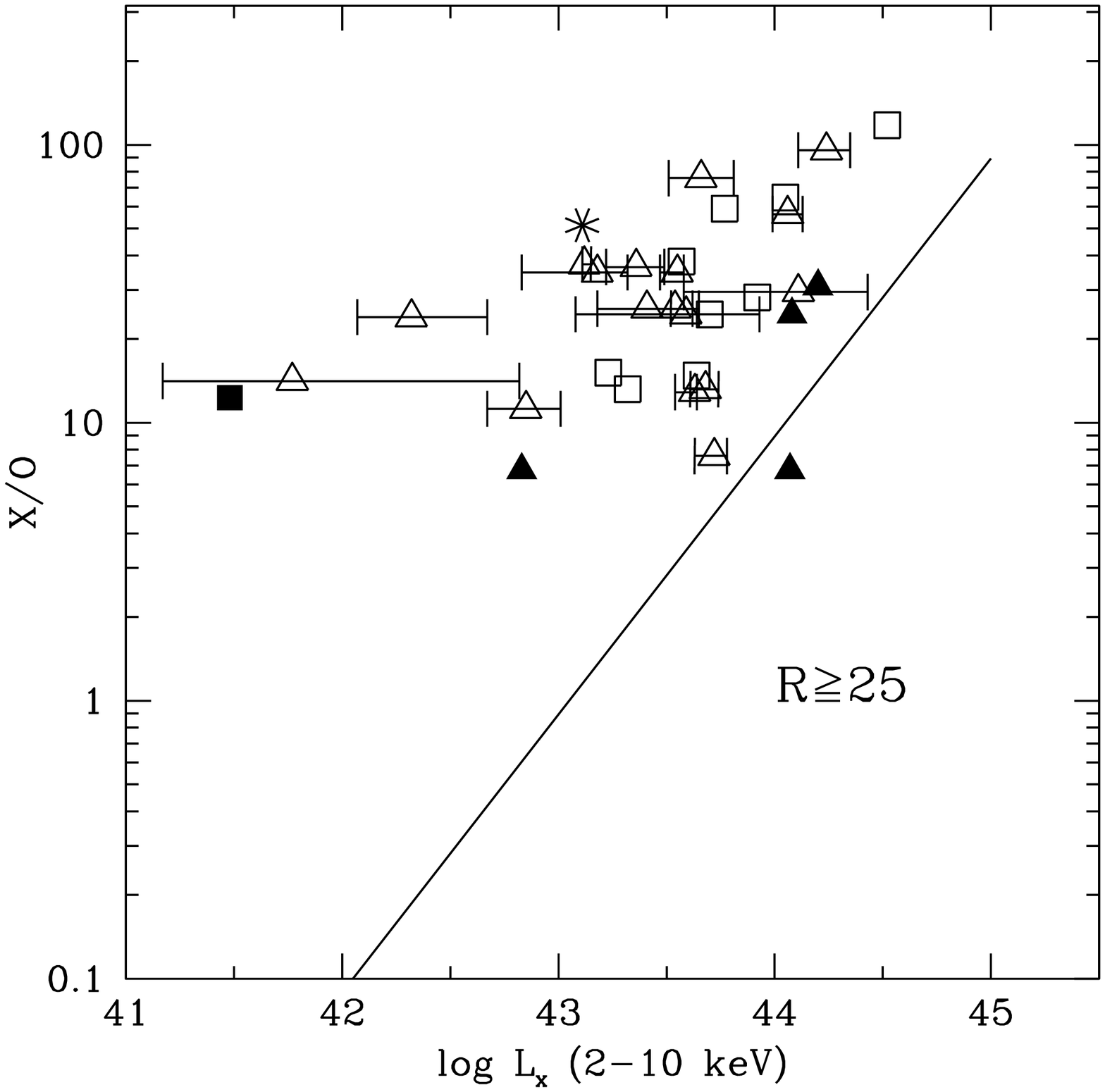}
\caption{The X--ray to optical flux ratio as a function of the 2-10 keV deabsorbed luminosity for the 61 sources 
with redshift
information (\cdfn\ = squares, \cdfs\ = triangles, filled symbols = spectroscopic redshift, open symbols = photometric
redshift, stars = redshift trough iron line). The error bars corresponds to the quoted uncertainties on the photo-z.
Left panel: optically "bright" sources (R$<$25). Right panel: optically faint objects (R$\geq$25). 
The extrapolation of the Fiore et al. (2003) relation is also reported
(continuous line). } 
\label{fig:nine}
\end{figure*}
A clear trend is evident in Fig.~\ref{fig:eight} suggesting
an increasing of the average column density towards high
redshifts. Such a correlation might well be due to a selection
effect. Obscured sources at low redshift could have been
missed simply because were too faint. At higher redshifts the
photoelectric cut--off is moved towards lower energies where 
the {\it Chandra} effective area is higher, favouring the 
detection of higher column densities.

In order to verify whether the observed trend is due to a selection 
effect, extensive Montecarlo simulations have been performed
using the  \ttt{fakeit} routine within {\tt XSPEC}.
A power--law spectrum with  $\Gamma$=1.8, normalized 
to encompass the observed range of 2--8 keV fluxes, is simulated 
for different values of the column density ($10^{22-23.5}$~cm$^{-2}$) 
over the redshift interval  $z$=0.1--3.5.
For each redshift the maximum \nh value for which
a source could be detected at the faintest flux of our 
sample has been computed. 
The results (Fig.~\ref{fig:eight}) indicate that obscured sources 
(\nh $\sim$ 10$^{23}$~cm$^{-2}$) at relatively low redshift ($z\simeq$1) 
would have been detected in the deep {\it Chandra} field.
 It should be noted that the dashed curve in Fig.~\ref{fig:eight}
would be shifted towards lower $N_H$ values, if the power law slope
is left free to vary in the range $\Gamma\simeq$ 1--1.4. 
However, also the observed $N_H$ value, with $\Gamma$ free to vary,
moves to lower values (see Table~5), leaving almost unchanged 
the bias estimated with $\Gamma$=1.8.

As a further check, we have simulated the expected 
distribution of counts in the very hard 4--7 keV 
band for the range of column densities and redshifts considered here.
The results indicate a change in the number of counts of $\sim$10--20\% (at 
most) at the highest column densities and lower redshift.
Overall, this counts distribution is consistent with the one observed in the 4--7 keV 
energy range in the stacked spectra of different 
redshift bins. Although we are confident that the present 
findings are not strongly biased against low--redshift absorbed sources
such a possibility cannot be completely ruled out.
Indeed, the simulations suggest that the sensitivity 
limit to absorption decreases, as expected, 
at low redshifts. As a result, for a given column density (\nh $\geq$ 10$^{22}$~cm$^{-2}$), faint sources
are more easily missed if they are at lower redshift. 

\section{Discussion}

The results of spectral analysis of a large sample of high X/O sources
in the {\it Chandra} deep fields leave no doubts about the 
presence of X--ray absorption. Similar results 
have been recently obtained for a sample of high X/O sources 
detected at brighter fluxes ($>$ 10$^{-14}$ \cgs\ ) 
in the {\tt HELLAS2XMM} survey (Perola et al. 2004; Comastri \& Fiore 2004). 

The shape of the stacked spectrum in both 
\cdfn\ and \cdfs\ and the residuals with respect to a single 
power--law fit are consistent with those expected by 
obscured sources spread over a range of redshifts.
The very hard average slope $\Gamma \simeq$ 0.9--1.0 in the \cdfs\ and 
$\Gamma \simeq$ 0.6--0.9 in the \cdfn\  is consistent with 
that obtained for a sample of hard (2--7 keV) and
ultrahard (4--7 keV) selected sources ($\Gamma \simeq$ 1.0--1.1) 
in the {\it Chandra} Groth Strip survey (Nandra et al. 2004).
Surprisingly, the fraction of the high X/O sources in the 
{\it Chandra} deep fields (about 23\%) is comparable to that (about 25\%) 
of hard spectrum objects in the Groth Strip sample.
The most straightforward explanation would imply that most 
of the hard sources in the Groth Strip survey have high 
X--ray to optical flux ratios. Such a possibility could be easily
tested thanks to the multiband imaging including HST/ACS observations.
It is also worth noting that the average spectrum of the sources in the 
Groth Strip survey has been computed with the hardness ratio technique 
and thus affected by larger uncertainties.

Most interesting, the average spectral shape of our sample 
as a function of the hard X--ray flux is approximately constant. 
Such a behaviour is significantly different from the 
well established trend observed in the same {\it Chandra} 
fields (Rosati et al. 2002; Alexander et al. 2001, 2003) without 
a specific selection on the X--ray to optical flux ratio.
Given that fainter X--ray sources are most likely at 
higher redshifts, the observed shape can be explained 
if more distant sources are, on average, more obscured.
Indeed, such a trend seems to be present 
for those sources (about half of the sample) for which
a redshift estimate has been considered. Though potentially 
interesting, such a correlation needs to be confirmed 
by a larger sample, with a larger fraction of spectroscopic identification, 
allowing to keep the observational biases under control. 

Unfortunately, the spectroscopic identification of obscured 
high X/O sources is already challenging the capabilities 
of large 8--10 m class telescopes.
An alternative method to multiband optical/near--infrared 
photometry and Fe K$\alpha$ line X--ray spectroscopy 
has been put forward by Fiore et al. (2003).
The X/O ratio and 2--10 keV luminosity of 
a large sample of spectroscopically identified hard X--ray 
selected sources follow a linear relation :
$\log L_{2 - 10} = log f(2 - 10 keV)/f(R) + 43.05$. 
The correlation holds for optically obscured sources (i.e., non broad--line AGN) 
and has been calibrated combining the 
optical and X--ray data of the {\tt HELLAS2XMM} survey
with well defined subsamples of identified sources 
in the deep {\it Chandra} fields at fluxes larger than 
3$\times 10^{-15}$~\cgs,
and optical magnitudes brighter than the R$\simeq$24--25.
Though characterized by a not--negligible dispersion (about 0.4 dex), 
this relation can be used to compute X--ray luminosities, 
and then redshifts of obscured sources, from the observed $X/O$ value.
The accuracy in the redshift estimate (``X--photo--z''; Fiore 2004)
is fairly good  [$\sigma(\Delta z / (1+z)) \simeq$ 0.2].

In order to check whether the Fiore et al. relation could be used
to infer redshifts for the unidentified sources in our sample, 
we have plotted the X/O values and luminosities for the 
61 sources\footnote{We did not include 1 broad line AGN because the 
Fiore et al. correlation does not hold for type 1 AGN.} 
in our sample with redshift information.  
The two panels of Fig. 9 show the unabsorbed 2--10 keV luminosity, 
calculated from the published 
2--8 keV counts and exposure times (Alexander et. al 2003) assuming 
a power--law model with $\Gamma$=1.8 plus Galactic absorption,
versus the X/O ratio for sources brighter and fainter of R=25, 
respectively. In each panel sources with spectroscopic, photometric 
and X--ray redshifts are plotted with different symbols and the Fiore et al. 
relation is also reported. While for the bright subsample there is a fairly
good agreement, though characterized by a significant dispersion, with 
the correlation, at fainter magnitudes high X/O sources have lower 
luminosities than expected.
Such a discrepancy, already noted by Bauer et al. (2004) 
and Barger et al. (2005), may indicate that the relative contribution 
of the host galaxy emission to the measured optical luminosity
at $R >$ 24--25 is different  than at brighter magnitudes. 
As a consequence some caution should be used when the Fiore et al. 
relation is extrapolated to optically faint objects. \\

The X/O vs. X--ray luminosity relation has been employed by 
Padovani et al. (2004) to search for type 2 quasars combining
deep {\it Chandra} exposure with HST/ACS imaging within the GOODS project. 
Almost by definition a large fraction of the candidate 
type 2 quasars (\nh $>$ 10$^{22}$~cm$^{-2}$, $L_X >$ 10$^{44}$~erg~s$^{-1}$) 
of the  Padovani et al. (2004) sample have X/O $>$ 1. 
Given that most of them have faint R--band magnitudes, the high
space density of type 2 quasars obtained by Padovani et al. (2004) 
should be considered as an upper limit.

Although the observed trend seems to indicate that the X/O vs. luminosity 
relation does not hold at $R >$ 25, it is important to note that
at faint optical magnitudes the probability to find by chance a galaxy 
in the X--ray error box increases dramatically (i.e. up to about 0.25--0.30
for R=26 and an error circle radius of 2 arcsec, without considering
source clustering). Furthermore, photo--z estimates 
often involve the determination of the magnitudes of the sources 
in images of very different quality (e.g., space versus ground
based telescopes) and, as a consequence, are affected by systematic errors
(see Zheng et al. 2004 for a discussion on the accuracy of 
redshift estimates in the \cdfs). 

\section{Conclusions}

The most important results obtained from the 
spectral analysis of a large sample of high X/O 
sources selected from deep {\it Chandra} fields 
can be summarized as follows:

\begin{itemize}

\item[$\bullet$] 
The average slope obtained by fitting the stacked spectra with a 
single power--law model in both \cdfn\ and \cdfs\ is extremely 
flat ($\Gamma \sim$1). The shape of the residuals strongly suggests that 
such a hard slope may well be the result of the superpositions
of sources with different intrinsic column densities over 
a broad range of redshifts.

\item[$\bullet$] 
The average slope of the stacked spectrum is almost independent
from the 2--8 keV flux.
The high X/O sources represent the most obscured component of the
X--ray background. Their spectra are harder ($\Gamma \lesssim$ 1) 
than any other class
of sources in the deep fields and also of the XRB spectrum.

\item[$\bullet$]
The redshift estimates, mainly obtained by multiband 
optical and near--infrared imaging, available for about half 
of the sample, allowed us to investigate the amount of intrinsic 
absorption in the stacked spectra for a few redshift ranges.
A trend of increasing absorption with redshift has been uncovered 
(N$_H = 10^{22-23.5}$~cm$^{-2}$).
While observational biases could not be completely ruled
out, extensive simulations suggest that the observed trend 
may well be real. Also, the rather constant hard spectrum 
over more than two orders of magnitude X--ray flux is consistent 
with such a behaviour. 

\item[$\bullet$]
Two new X--ray redshifts, obtained from the K$\alpha$ iron line, 
have been discovered. One object satisfy the criterion of 
an X--ray type 2 QSO.

\item[$\bullet$]
Though the X/O ratio is a fairly good estimator of the source
luminosity and redshift (Fiore et al. 2003), at faint optical
magnitudes (R$>$ 25) a departure from the X/O vs. luminosity relation 
is emerging.  
\end{itemize}

A large sample of high X/O sources at relatively bright X--ray 
fluxes will be obtained in the next few years with the COSMOS 
survey: a contiguous 2 square degree area observed with 
ACS/HST where, along with medium--deep
XMM exposures, a large body of multiwavelength data is available.
The full exploitation of this unique database will 
allow us to quantitatively estimate the redshift and absorption 
distribution of bright high X/O sources.

\section{Acknowledgments}
It is a pleasure to thank Piero Ranalli for help
in data reduction, Fabrizio Fiore, Roberto Gilli, Guido Risaliti, 
Paolo Tozzi and Cristian Vignali for useful discussion. An anonymous referee
is also thanked for a fast report which allow us to improve the 
presentation and discussion of the results.
The authors acknowledge partial support by
the MIUR grant COFIN--03--02--23 and by the INAF grant 270/2003.

\end{document}